\documentclass[fleqn,twoside,twocolumn,nofootinbib]{revtex4} % Specifies the document class %,unsortedaddress
\usepackage[sec]{ujp} % \usepackage[cyr]{ujp} for cyrillic, \usepackage[web]{ujp} for web
\begin{document}
\title[RADIATIVE TRANSITIONS BETWEEN THE AUTOIONIZING STATES]%колонтитул
{RADIATIVE~ TRANSITIONS~ BETWEEN THE \boldmath${4d^{10}5p^2(^3P_j
)nl}$ AND ${4d^{10}5s5p(^3P_1^o )nl}$
AUTOIONIZING STATES OF In ATOM IN COLLISIONS OF SLOW ELECTRONS WITH In$^{+}$ IONS}%
\author{G.M. GOMONAI, A.I. IMRE, E.V. OVCHARENKO, YU.I. HUTYCH}%1 автор
\affiliation{Institute of Electron Physics, Nat. Acad. of Sci. of Ukraine}%институт
\address{21, Universytets'ka Str., Uzhgorod 88017}%адрес
\email{dep@mail.uzhgorod.ua}%e-mail

\avtorcol{G.M. GOMONAI, A.I. IMRE, E.V. OVCHARENKO, Yu.I. HUTYCH}%%колонтитул
\udk{539.186.1}  \razd{\seciii}

\setcounter{page}{957}%
\maketitle

\begin{abstract}
The~ radiative~ transitions~ between~ the~ $4d^{10}5p^2(^3P_{0,1,2}
)nl$ and \mbox{$4d^{10}5s5p({}^3P_1^o )nl$~autoionizing} states of
an indium atom that represent dielectronic satellites of the
$4d^{10}5p^2\;{ }^3P_0 \to 4d^{10}5s5p\;{ }^3P_1$ ~($\lambda
=171.7$~nm), $4d^{10}5p^2\;{ }^3P_1 \to 4d^{10}5s5p\;{ }^3P_1$
~($\lambda =166.7$~nm), and $4d^{10}5p^2\;{ }^3P_2 \to
4d^{10}5s5p\;{ }^3P_1$ ~($\lambda =160.7$~nm) spectral lines of an
In$^{+}$ ion have been observed for the first time. The energy
dependences of the effective excitation cross sections of
dielectronic satellites, as well as near-threshold regions of these
spectral lines, were investigated in the range of electron energies
9$\div $15~eV with the help of the spectroscopic method using
crossed beams of electrons and In$^{+}$ ions. The absolute
excitation cross sections of dielectronic satellites amount to
$(0.7\div 2)\times 10^{-16}$~cm$^{2 }$ and are of the same order of
magnitude as the effective excitation cross sections of the
corresponding ionic lines. It is found that a considerable increase
of the probability of radiation decay of the
$4d^{10}5p^2(^3P_{0,1,2} )nl$ autoionizing states of the In atom is
related to strong relativistic and correlation effects, in
particular to the configuration interaction of the $5p^2nl$ levels
both with one another and with the levels of the $5s5dnl$ and
$4d^95s^25p^2$ configurations.
\end{abstract}

\section{Introduction}

Progress in many fields of science and technology depends on the
knowledge about quantitative and qualitative characteristics and
mechanisms of the processes accompanying electron-ion collisions.
Data on these processes are important for the successful development
of such areas of the modern science as plasma physics, astrophysics,
physics of nuclear reactions with heavy ions, laser and analytical
technology, thermonuclear energetics, quantum chemistry, {\it etc.}
[1--3]. Elementary processes of electron-ion collisions also clearly
manifest themselves in the phenomena taking place in the upper
atmosphere of the Earth and other planets. It is worth noting that
spectral lines of heavy many-electron elements have been recently
observed in spectra of stars and in the interstellar
space~[4].\looseness=1

The results of experimental and theoretical researches performed
during recent years in leading scientific groups of the world
testify to the complicated mechanism of inelastic processes running
in the case of collisions between slow electrons and ions [5,6]. It
is mainly explained by resonance effects related to the formation
and decay of autoionizing states (AIS) of the ``electron+ion''
system. In the case of electron-ion collisions, the long-range
Coulomb field of an ion not compensated by electrons results in the
infinite number of AISs converging to each ion level. They decay
both via the electron channel (radiationless decay accompanied by
the ejection of an electron), which results in a complicated
resonance structure of the scattering cross sections and a
considerable addition to the effective cross section of direct
excitation, and via the radiation channel in the process of
dielectronic recombination (DR) of the ion. As is known [7,8], in
addition to the DR process proper (which is in many cases
determinative for the ionization equilibrium of plasma), DR of ions
also manifests itself in the form of satellites of the resonance and
other lines of an ion appearing in the case of radiative transitions
from AISs (so-called dielectronic satellites). They are present in
spectra of recombining plasma (for example in spectra of solar
flares, laser plasma, plasma of tokamaks, {\it etc.}). The
wavelengths of dielectronic satellites are close to those of the
corresponding spectral line of an ion and very suitable for the
diagnostics of laboratory and astrophysical plasmas, as the ratio
between~~ the intensities of satellites and resonance lines
essentially depends on the temperature~[9].\looseness=1

%Fig. 1
\begin{figure}% figure* for wide figure, [h] [!] to change the placement
\includegraphics[width=\column]{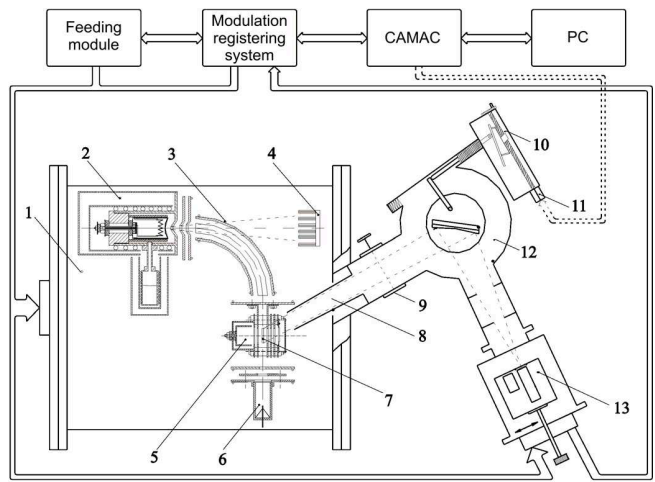}
\caption{Diagram of the set-up: {\it 1} -- collision chamber, {\it
2} -- ion source, {\it 3} -- ion selector, {\it 4} -- cooled atom
trap, {\it 5} -- electron gun, {\it 6} -- ion collector, {\it 7} --
crossing region of beams, {\it 8} -- radiation from the collision
area, {\it 9} -- vacuum shutter, {\it 10} -- control mechanism of
the grating rotation, {\it 11} -- step motor, {\it 12} -- vacuum
monochromator, {\it 13} -- radiation detector unit}
\end{figure}

A special kind of this type of dielectronic satellites is presented
by those radiating in the case of radiative transitions between
AISs. This radiation channel of decay is realized for the AISs that
converge to ion levels located above resonance (lowest) ones. It is
worth noting that, during almost three decades of experimental
investigations of DR, the radiation observed in experiments
corresponded to transitions, for which only the upper level was
autoionizing, whereas the lower level represented one of the bound
levels of a neutral atom (or an ion with charge multiplicity lower
by unity in the case of multiply charged ions) [2, 5]. Radiative
transitions between the $4d^9({ }^2D_{5/2,3/2} )5s^2nl$ and
$4d^{10}5p({ }^2P_{1/2,3/2}^o )nl$ AISs in the process of DR in the
case of electron-ion collisions were for the first time observed in
the course of investigations of the near-threshold regions of the
effective excitation cross sections of the $(4d^95s^2{ }^2D_{3/2}
\to 4d^{10}5p{ }^2P_{1/2,3/2}^o )$ laser transitions of a Cd$^{+}$
ion [10] (there were also obtained the energy dependences of their
effective cross sections). The results of this research have
demonstrated that the effective excitation cross sections of
dielectronic satellites amount to $\sim 10^{-17}{~}$cm$^{2}$ and are
comparable to the effective excitation cross sections of the
corresponding spectral lines.\looseness=1

Investigations of the electronic excitation of an In$^{+}$ ion are
first of all interesting from the viewpoint of atomic physics, as it
represents an atomic system with completed valence $(5s^2)$ and
subvalence $(4d^{10})$ shells. Moreover, it is characterized by
strong correlations both inside the shells and between them and the
effective simultaneous excitation (except for one of the $s$- or
$d$-electrons) of two $s$-electrons, which considerably complicates
the spectrum of this ion.

In this work, we present the results of the spectroscopic analysis
of the excitation of the radiative transitions between the
$4d^{10}5p^2(^3P_{0,1,2} )nl$ and $4d^{10}5s5p({ }^3P_1^o )nl$ AISs
of an indium atom in the case of collisions between slow electrons
and In$^{+}$ ions.

\section{Experimental Apparatus}

The experiment was performed using the spectroscopic method under
the crossing of the electron and In$^{+}$ ion beams at the right
angle. As one can see from the diagram of the set-up presented in
Fig.~1, it consists of a high-vacuum collision chamber that contains
an ion source, ion selector, electron gun, vacuum monochromator with
a mechanism allowing one to control the rotation angle of the
diffraction grating, unit of radiation detectors, vacuum pumping
system, power supply unit, modulation system for registration of the
investigated radiation, CAMAC crate, and personal computer for the
processing and storage of data. Experiments with indium metal make
strict demands for the construction of the ion source, breakdown
protection of the isolators of the ion-optical lens system, and the
choice of optimal parameters of the experiment. The basic units of
the set-up are described in [11,12] in detail. The present work
includes only the aspects most important for performing these
precision measurements.

A number of technical improvements of the construction of the ion
source, as well as the optimal mode of its operation, allowed us to
obtain a stable beam of In$^{+}$ ions (mainly in the ground-state) with a
cross section of $2\times 2$~mm$^{2}$ and a current of $I_i =2\times
10^{-6}$ А~ at an energy of 700 eV. The low-energy three-anode gun
formed a strip-like electron beam with a cross section of $1\times 8$
mm$^{2}$ in the range 9$\div $15 eV with a current equal to $I_e
=(1\div 1.2)\times 10^{-4}$~А.

As is known, under real conditions of studying the electron-ion
collisions, one registers superweak signals from the investigated
processes, which is related, first of all, to low ($<10^7$~cm$^{-3})$
concentrations of the interacting particles (which are five orders of
magnitude lower than the concentrations of atoms in similar experiments
in the case of electron-atom collisions). That is why, in order to
obtain a proper level of the useful signal, one should deal with
rather high values of $I_e$. In turn, this results in the worsening of
the energy homogeneity of the electron beam. Whence it follows that, adjusting the electron gun, it is necessary
to find a reasonable compromise in the matching of the electron
current $I_e$ and the maintenance of a sufficient energy homogeneity
of the electron beam. Thus, the energy homogeneity of the electron
beam in our experiments amounted to $\Delta E_{1/2} \sim 0.8$~eV.

Spectral decomposition of the radiation was realized with the help
of a 70-degree Seya--Namioka vacuum monochromator with a concave
toroidal grating (1200~lines per mm) and the reciprocal linear
dispersion $\partial \lambda /\partial l\sim 1.7$~nm/mm. The
radiation was detected by means of a solar blind photomultiplier.
The time of data storage at each experimental point amounted to
2000~s, whereas the magnitude of the useful signal was equal to
$0.5\div 0.2$~ pulses per second at a signal/background ratio from
1/10 to 1/30.

In order to obtain the absolute cross section of electron excitation
of dielectronic satellites, it was necessary to determine the
spectral sensitivity of the experimental apparatus $\eta _{\lambda}$
[2]. For this purpose, the intensity of the investigated radiation
was to be thoroughly compared to the absolute standard. As the
studied spectral lines lie in the wavelength range 100--200 nm (that
is, in the ultraviolet spectral region), the most accurate
calibrated source in this region is the synchrotron radiation of
electron accelerators. As we had no possibility to use a
synchrotron, the spectral sensitivity of the set-up was determined
with the help of the technique, where the calibrated radiation
source was presented by one of the gases excited by the electron
impact. This technique was described in [13] in detail, and its
essence consists in that, under constant measurement conditions, the
spectral sensitivity of a set-up $\eta _{\lambda}$ is a function of
only the effective excitation cross section of the spectral line
$\sigma $. Thus, knowing the electron excitation cross sections of
several spectral lines of the given gas, we can determine the
relative spectral sensitivity of the experimental set-up. In our
case, the discrete values of the function $\eta _{\lambda
}=f(\lambda )$ are determined according to the formula
%1
\begin{equation}
\eta_{\lambda _i} = \frac{C_{1} \sigma_{i}}{C_{i} \sigma _{1}}
\eta_{\lambda _{1}} ,
\end{equation}
where $\eta _{\lambda _i} $ and $\eta _{\lambda _1} $ denote the
sensitivities for two spectral lines with the wavelengths $\lambda
_i$ and $\lambda _1$, respectively, $C_{i}$ and $C_1$ are the
signals measured at these wavelengths, and $\sigma _{i}$ and $\sigma
_1$ represent the corresponding effective excitation cross sections.
Index $1$ indicates the spectral line, for which $\eta _{\lambda }$
was accepted to be equal to unity, and the number of values taken on
by $i$ is equal to the number of lines with known excitation cross
sections.

Based on the above-stated considerations, we determined the relative
spectral sensitivity of the experimental set-up with the help of the
electron-impact excitation cross sections of nitrogen spectral lines
[14]. Our measurements were carried out at an electron energy of
100~eV and a pressure in the collision chamber of $10^{-5}$~Torr.
The absolute spectral sensitivity of the set-up in the wavelength
range 110$\div $180~nm was determined with an error not exceeding
30~{\%} using, as a reference, the absolute electron-impact
excitation cross section of the resonance line of an In$^{+}$ ion at
the wavelength $\lambda =158.6$~nm [11] obtained by normalizing the
experimental data at an energy of 300~eV to the results of
theoretical calculations by the semiempiric Van Regemorter formula.

\section {Results and Their Discussion}

An In$^{+}$ ion is isoelectronic with respect to a cadmium atom that
obeys the general regularities of spectra with completed valence
$5s^2$ and subvalence $4d^{10}$ shells. The ground state of the
In$^{+}$ ion has the $4d^{10}5s^2\;{ }^1S_0 $ configuration. A
simultaneous excitation of two valence $5s^2$ electrons results in
the formation of additional ion states along with the common ones.
Those are the so-called ``shifted terms'' [15], for which the
radiative decay to the ground state is forbidden according to the
selection rules. That is why they can combine only with each other
or with common terms. The positions of the $^3P_{0,1,2}$, $^1D_2 $,
and $^1S_0 $ shifted terms of the $5p^2$ configuration for the
In$^{+}$ ion are known from the literature [16] (see Fig. 2).
According to the selection rules for shifted terms [15] (if $\Delta
l_i =\pm 1$ for one electron, then $\Delta l_i =0,\;\pm 2$ for the
other one; moreover, $\Delta J=0,\;\pm 1$, except for the cases $J_1
=0\to J_2 =0$), the most probable decay channel of the levels of the
$5p^2$ configuration is their radiative decay into the
$5s5p\;(^3P_j^o,\;^1P_1^o)$ resonance levels of the In$^{+}$ ion.

%Fig. 2
\begin{figure}% figure* for wide figure, [h] [!] to change the placement
\includegraphics[width=\column]{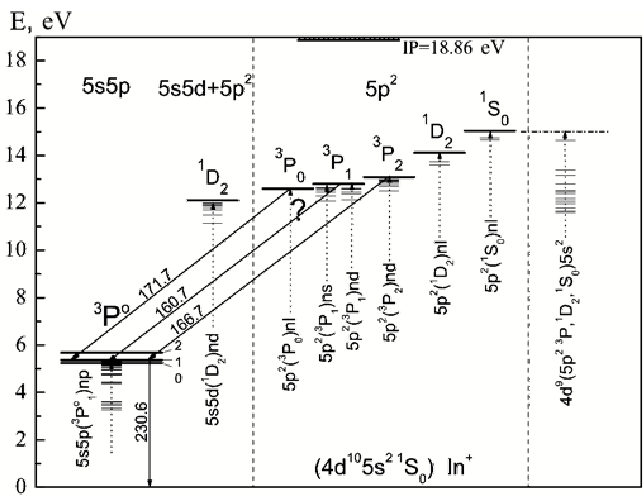}
\caption{Diagram of the lower levels of an In$^+$ ion and AISs
converging to them}
\end{figure}

In the course of studying the electron excitation of radiative
transitions from the $4d^{10}5p^2\;{ }^3P_{0,1,2} $ shifted terms of
the In$^{+}$ ion [12] according to the reaction
\[e+{\rm In}^{+}(4d^{10}5s^2){ }^1S_0 \to e_1 +{\rm In}^{+\ast }(4d^{10}5p^2){
}^3P_{0,1,2}\to \]
%2
\begin{equation}
\to e_1 +{\rm In}^{+\ast }(4d^{10}5s5p){ }^3P_1^o +h\nu _{1,2,3}
\end{equation}

\noindent ($\lambda _1 =171.7$ nm, $\lambda _2 =167.7$ nm, and
$\lambda _3 =160.7$ nm), we revealed a radiation lower than the
excitation thresholds in the wavelength interval $\lambda _i \pm
2.0$~nm with respect to each line. In order to interpret the nature
of this radiation, we performed the precision investigations of the
near-threshold regions of the energy dependences of the electron
excitation of the indicated spectral lines.

Figure~3 presents the investigation results, the confidence interval
indicated at each point with the probability 68~{\%}, and the
threshold excitation energies of the shifted terms of the In$^{+}$
ion [17]. The mean square error of relative measurements did not
exceed $\pm 20$~{\%} in the whole investigated energy interval. The
accuracy of the energy scale determination amounted to $\pm 0.1$~eV.
The absolute values of the measured excitation functions were
obtained to within $\pm$30\%.

As one can see from Fig.~3, the energy dependences in the region
below the excitation thresholds of the $5p^2\;{ }^3P_{0,1,2} $
levels (9$\div $12~eV) are characterized by resonance peculiarities.
Comparing their energy positions to the excitation energies of the
$5p^2({ }^3P_j )nl$ autoionizing states of the In atom with regard
for the fact that these resonance peculiarities result from
the radiation of satellite lines lying in the experimentally
observed wavelength interval ($\lambda _i \pm 2.0$~nm), we concluded
that they are related to the excitation of the radiative transitions
between the $5p^2({ }^3P_j )nl$ [16] and $5s5p({ }^3P_1^o )nl$ [18]
AISs. Moreover, the most probable mechanism of their excitation is
the DR process:
%3
\[e+{\rm In}^+(4d^{10}5s^2)\;{ }^1S_0 \to {\rm In}^{\ast \ast }\left[ {4d^{10}5p^2({}^3P_{0,1,2} )nl}\right]\to \]
\begin{equation}
\to {\rm In}^{\ast \ast }\left[ {4d^{10}5s5p({ }^3P_1^o )nl}
\right]+h\nu _n ,
\end{equation}

%Fig. 3
\begin{figure}% figure* for wide figure, [h] [!] to change the placement
\includegraphics[width=6.7cm]{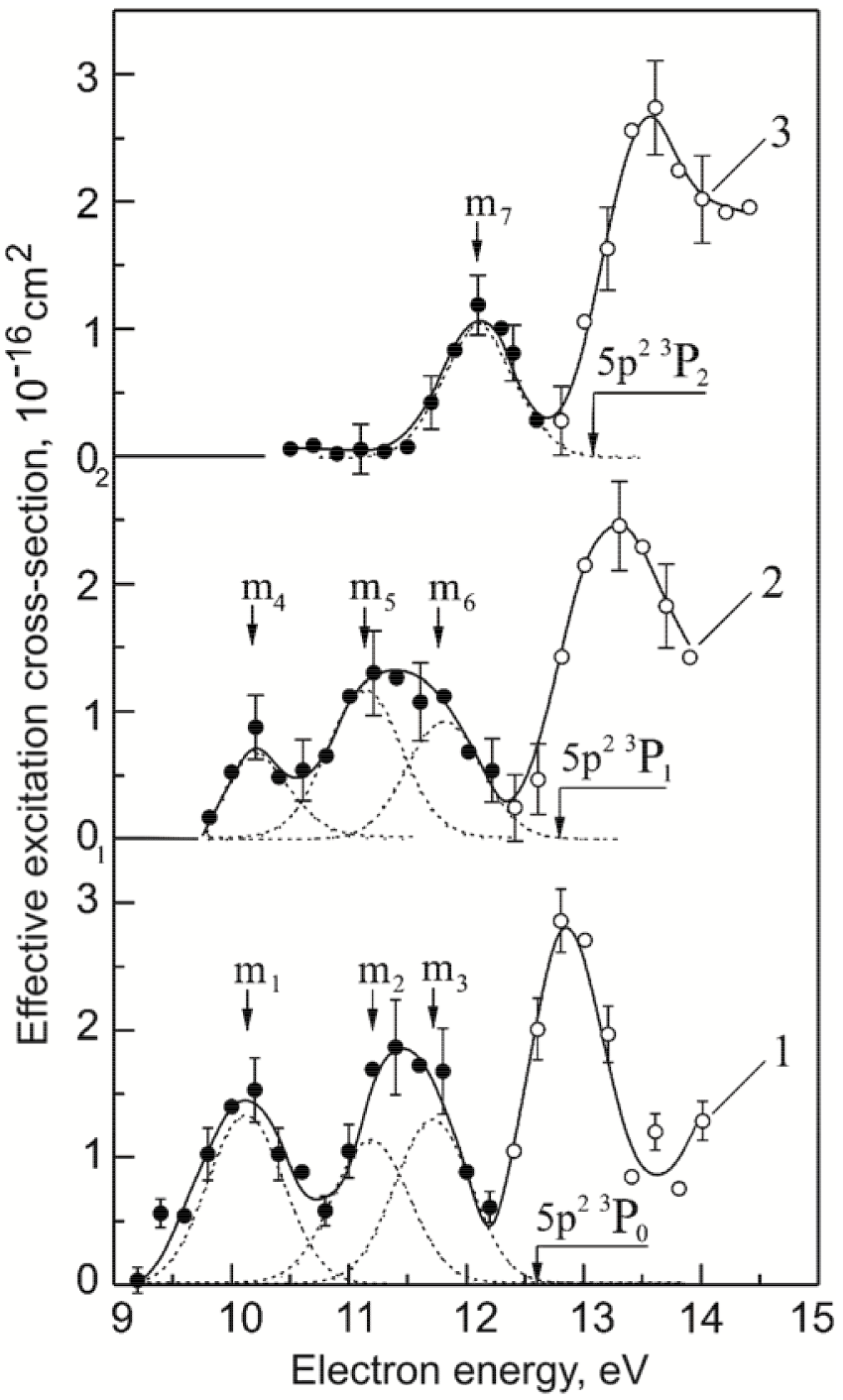}
\caption{Energy dependences of the effective excitation cross
sections of dielectronic satellites of the spectral lines $\lambda=
171.7$~nm ({\it 1}), $\lambda =166.7$~nm ({\it 2}), $\lambda=
160.7$~nm ({\it 3}) of an In$^+$ ion}
\end{figure}

\noindent where $\lambda _1 =171.7\pm 2.0$~nm; $\lambda _2 =167.7\pm
2.0$~nm; $\lambda _3 =160.7\pm 2.0$~nm, i.e. they represent
dielectronic satellites of the studied spectral lines. It is worth
noting that, in this case, the radiation stabilization of AISs is
realized to the AISs of the $5s5pnl$ configuration (rather than to
the bound states of the atom), whose convergence limit is presented
by the $5s5p\;{ }^3P_1^o $ resonance level of the In$^+$ ion.
Moreover, the dielectronic satellites are most probably formed with
participation of only those AISs of the $4d^{10}5p^2nl$
configuration, whose energies are lower than that of the $5p^2\;{
}^3P_0 $ level. Above the excitation threshold of this level, there
exists a more energy-efficient electron channel of decay of such
AISs into levels of the $5p^2$ configuration of this ion.

Due to the lack of data on energy positions of the AISs of the
indium atom, a clear interpretation of the revealed structure is
complicated. For example, among all possible AISs converging to the
levels of the $5p^2$ configuration, only the energy positions of the
AISs of the $5p^2({ }^3P_1 )ns$, $5p^2({ }^3P_1 )nd$, $5p^2({ }^3P_2
)nd$, $5p^2({ }^1D_2 )nl$, and $5p^2({ }^1S_0 )nl$ configurations
are present in the literature [16]. It is strange that there are no
data on the $5p^2nl$ AISs, whose convergence limit is the $^3P_0$
level, though such AISs for Ga$^{+}$ and Tl$^{+}$ ions are known
[16]. However, for this energy region (below the $5p^{2}\;{ }^3P_0 $
level), there were observed AISs identified in [16] as those of the
$5p^2({ }^1D_2 )nd$ configuration. That is, it was considered that
the $5p^{2}\;{ }^1D_2$ level in In$^{+}$ is located below the
triplet $5p^2\;{}^3P_j$ levels. The investigations performed in [17]
gave clear evidences of the fact that there exists a strong
configuration interaction between the $5p^{2}\;{ }^1D_2 $ and
$5s5d\;{ }^1D_2 $ levels, both of them divide between the $5p^{2\;}$
and $5s5d\;$ configurations. Moreover, only the latter configuration
can be ascribed to the lower level. In addition, there takes place
the configuration mixing among all the shifted levels of the
$5p^{2\;}$ configuration, as well as among the shifted and ordinary
levels. It follows from here that the AISs converging to the
above-mentioned levels are also configurationally mixed, which
considerably complicates, on the one hand, the identification of the
revealed structural peculiarities, but, on the other hand,
represents a strong argument of the fact that such a configuration
mixing of the levels substantially increases the probability of
radiation decay of the AISs.\looseness=1

It is worth noting that the obtained electron excitation functions
are a superposition of the true electron excitation functions of the
investigated transitions, and the energy homogeneity of the electron
beam $\Delta E_{1/2} \sim 0.8$~eV. That is why, in order to study
the structural peculiarities of the energy dependences below the
electron excitation thresholds of the $(4d^{10}5p^2\;{ }^3P_j \to
4d^{10}5s5p\;{}^3P_1^o )$ radiative transitions, we carried out the
procedure of decomposition of the measurement results into separate
components taking the magnitude of the instrument function into
account by the procedure presented in~[19].\looseness=1

With regard for the above-stated considerations, we may suppose
that the structure of the studied energy dependences of the
effective excitation cross sections below an energy of 10.8~eV
(maxima $m_1 $ and $m_4)$ is caused by the radiative transitions
between the lowest ($n = 6$--8) $4d^{10}5p^2({ }^3P_{0,1} )np$ and
$4d^{10}5s5p({ }^3P_1^o )np$ AIS configurations. As for the
structural peculiarities in the energy range from 10.8 to 12.6~eV
(maxima $m_2$, $m_3$, $m_5$, $m_6$, and $m_7$), they are resulted probably
from the radiative transitions between the lowest
$4d^{10}5p^2({ }^3P_{0,1,2} )ns,np,nd$ and $4d^{10}5s5p({}^3P_1^o
)ns,np,nd$ AIS configurations. It is worth noting that, first, the
decay of these AISs is of random nature. Second, the AISs of the
$4d^95s^25p^2$ configuration are also efficiently excited in the
investigated region of electron energies [16], which can
considerably influence the above-mentioned mechanisms of excitation
of the investigated dielectronic satellites.

\section {Conclusions}

The analysis of the  energy dependences of the effective excitation
cross sections of dielectronic satellites of the spectral lines
($4d^{10}5p^2\;{ }^3P_j \to 4d^{10}5s5p\;{ }^3P_1^o $) of the
In$^{+}$ ion has demonstrated that, along with the excitation of
dielectronic satellites of the resonance lines corresponding to the
radiative transitions between AISs and bound atomic states [20],
there also takes place the effective excitation ($\sigma \sim
10^{-16}$~cm$^{2}$) of those related to the radiative transitions
between the $4d^{10}5p^2(^3P_{0,1,2} )nl$ and $4d^{10}5s5p({
}^3P_1^o )nl$ AISs of the indium atom. Moreover, the radiative
transitions ($4d^{10}5p^2\;{ }^3P_j \to 4d^{10}5s5p\;{ }^3P_1^o $)
(or radiative cascades) represent an important factor of the
population of excited atomic states and, therefore, one of the basic
mechanisms of formation of the intensities of the spectral ion
lines, including the dielectronic satellites of resonance lines.

The radiative transitions between the AISs initiated by DR of an ion that
were chosen as the investigation object, as well as the obtained
results, are interesting from the viewpoint of general physics
and are important for the practical solution of a number of applied problems
of plasma physics and controlled thermonuclear fusion. Among the
latter, we mention the problem of ionization balance, the development of
radiation models for plasma diagnostics, {\it etc.}. As DR represents
an important element of the ionization balance and an error in its
determination can substantially influence the result, a correct
allowance for radiative transitions between AISs, whose main
excitation mechanism is DR, is a necessary condition of the
establishment of the proper relation between the observed
intensities of spectral lines and plasma parameters, which forms a
basis of its spectroscopic diagnostics.

Though the effective DR cross sections (or reaction rates) have been
theoretically investigated for decades, the used theoretical methods
give an error from 50{\%} to a factor of 2 [5]. In this case, the
calculated constants are considerably lower than the experimentally
observed values. The allowance for such additional factors as
relativistic effects and the influence of the external electric
field on the DR behavior allowed one to approach theoretical data to
experimental ones. But in spite of the substantial mathematical
apparatus used in theoretical works, a large discrepancy between
theory and experiment makes one think that the theoretical
calculations do not take into account some fundamental peculiarity
of DR. The analysis of the results of our researches gives grounds
to state that one of such peculiarities is the radiative cascade
from AISs to lower AISs.%\looseness=1

In addition to a substantial improvement of the experimental
conditions (in particular, the energy homogeneity of the electron
beam), a more solid analysis of the manifestation of AISs in
inelastic collisions of slow electrons with such complex
many-electron system as In$^{+}$ ion requires theoretical
calculations with regard for the radiation decay of AISs. This
especially concerns the energy region near the excitation thresholds
of spectral lines, where the resonance contribution of AISs is
dominant, which will allow one to solve a number of problems related
to the competition between the processes of autoionization and radiation.

\end{document}